\documentclass[aps,prb,twocolumn,amsmath,amssymb,superscriptaddress,floatfix]{revtex4}
\usepackage{graphicx}
\usepackage{bm}
\usepackage{amssymb}
\usepackage{color}
\usepackage{amsmath}
\usepackage{multirow}
\newcommand{\be}{\begin{equation}}
\newcommand{\ee}{\end{equation}}
\newcommand{\beqn}{\begin{eqnarray}}
\newcommand{\eeqn}{\end{eqnarray}}

\begin{document}
\title{Excess entropy and central charge of the two-dimensional random-bond Potts model in the large-$Q$ limit}
\author{Istv\'an A. Kov\'acs}
\email{kovacs.istvan@wigner.mta.hu}
\affiliation{Wigner Research Centre, Institute for Solid State Physics and Optics, H-1525 Budapest, P.O.Box 49, Hungary}
\affiliation{Institute of Theoretical Physics, Szeged University, H-6720 Szeged, Hungary}
\author{Jean-Christian Angl\`es d'Auriac}
\affiliation{Institut N\'eel-MCBT CNRS, B. P. 166, F-38042 Grenoble, France}
\author{Ferenc Igl\'oi}
\email{igloi.ferenc@wigner.mta.hu}
\affiliation{Wigner Research Centre, Institute for Solid State Physics and Optics, H-1525 Budapest, P.O.Box 49, Hungary}
\affiliation{Institute of Theoretical Physics, Szeged University, H-6720 Szeged, Hungary}

\date{\today}

\begin{abstract}
We consider the random-bond Potts model in the large-$Q$ limit and calculate the excess entropy, $S_{\Gamma}$, of a
contour, $\Gamma$, which is given by the mean number of Fortuin-Kasteleyn clusters which are crossed by $\Gamma$.
In two dimensions $S_{\Gamma}$ is proportional to the length of $\Gamma$, to which - at the critical point - there
are universal logarithmic corrections due to corners. These are calculated by applying techniques of conformal field theory and
compared with the results of large scale numerical calculations. The central charge of the model is obtained from
the corner contributions to the excess entropy and independently from the finite-size correction of the free-energy as:
$\lim_{Q \to \infty}c(Q)/\ln Q =0.74(2)$, close to previous estimates calculated at finite values of $Q$. 
\end{abstract}

\maketitle

\section{Introduction}
Entropy represents a fundamental concept in different domains of science, such as in information theory\cite{shannon},
in quantum systems\cite{entanglement_entropy}
and in classical statistical mechanics. In quantum systems the entanglement entropy turned out a very important indicator
of new and exotic phases and quantum phase transitions\cite{entanglement_entropy,holzhey,vidal,calabrese_cardy}.
Its analogue in information theory is the mutual information
and in classical statistical mechanics the excess entropy. Also in classical systems the excess entropy can be
defined as mutual information, where the probabilities are given
as Boltzmann weights\cite{wilms,lau_grassberger,mutual_classic}.
%=Z(\beta)^{-1}e^{-\beta E_i}$, where $Z(\beta)=\sum_i e^{-\beta E_i}$ is the partition function
%and $\beta=1/T$ is the inverse temperature.
In these calculations the system is divided in
two (or more) parts, and the excess entropy is basically associated with the interface separating the subsystems. Due
to this, the excess entropy is proportional to the surface of the interface, which is called the area law.
In critical systems, however, usually there are universal corrections to the area law, which are logarithmic in
the linear extent of the interface. In conformally invariant systems, such as in
one-dimensional quantum and in two-dimensional classical models the central charge of the conformal field theory can be
deduced from the size-dependence of the critical entanglement entropy\cite{entanglement_entropy,holzhey,vidal,calabrese_cardy}
and mutual information\cite{mutual_classic}, respectively.

Concerning classical statistical mechanics most of the studies described in the previous paragraph are performed on non-random systems.
It is known, however, that systems with quenched disorder are also conformally invariant, provided the properties of
averaged quantities (magnetization, correlation function, etc.) are concerned. Therefore it is of interest to study
the scaling properties of the excess entropy in random systems, too. For this purpose we consider the two-dimensional
$Q$-state Potts model\cite{Wu} in the presence of bond disorder\cite{Cardy99}. In two dimensions (2D) the phase transition in the
random bond Potts model
(RBPM) is of second order\cite{imry,aizenman_wehr,hui} for any
value of $Q$, even in the limit $Q \to \infty$. The critical behavior of this model has been studied by different
methods\cite{cardyj,jcardy,picco,palagyi,chatelain}, in particular the critical exponents and the central charge has been calculated, mainly for $Q>4$, in which case
the phase transition in the non-random model is of first order\cite{baxter}. Special attention has been paid to
the model in the large-$Q$ limit, in which case the critical parameters are found
to be smooth function of $1/\ln Q$\cite{jacobsen}.
For example from the numerical data calculated at large, but finite values of $Q$ the central charge is conjectured to be\cite{jacobsen}
$\lim_{Q \to \infty}c(Q)/\ln Q\equiv c' =1/(2\ln2)=0.7213$.

Later, it has been shown\cite{juhasz} that the model can be studied directly at the limiting value $Q \to \infty$,
when in the random cluster representation\cite{Fortuin-Kasteleyn}
the partition function of the model is dominated by one term, the so called \textit{optimal graph}. This means that
thermal fluctuations are negligible compared to disorder fluctuations, thus the critical behavior of the system is controlled
by a so called infinite disorder fixed point\cite{im}. The optimal graph of the RBPM has been calculated
by a combinatorial optimization method\cite{dips}, which provides the exact value of the partition function for a given sample, i.e.
for a given realization of the disorder. From the numerical data exact values of the critical exponents are
conjectured\cite{di2003,mdi2004} through an expected relation with the exactly known infinite disorder fixed point
of the random transverse-field Ising chain\cite{fisher}:
$x_m=(5-\sqrt{5})/4$ (bulk magnetization), $x_m^s=0.5$ (surface magnetization) and $\nu=1$ (correlation length).

In this paper we study the scaling properties of the excess entropy in the RBPM in the large-$Q$ limit. For this we consider a
subset of bonds, $\Gamma$, and calculate the corresponding excess entropy, $S_{\Gamma}$. If $\Gamma$ is a closed loop,
separating a subsystem, ${\cal A}$, from the rest of the system, ${\cal B}$, then $S_{\Gamma}$ is the mutual entropy
$S_{\Gamma}=S_{\cal A}+S_{\cal B}-S_{{\cal A} \cup {\cal B}}$. In the following section we show, that 
in the random cluster representation $S_{\Gamma}$ is simply given by the mean number of clusters in the optimal
sets which are crossed by $\Gamma$. This type of problem has already been considered by two of us in the case of the non-random
Potts model both for $Q=1$, representing percolation\cite{kovacs,kovacs_3d} and for general values of $Q \le 4$\cite{kovacs14}.
Repeating the reasoning applied
in these papers we show that the dominant term of $S_{\Gamma}$ represents the area law to which there are logarithmic
corrections at the critical point due
to corners and these are calculated by conformal techniques.
The analytical conformal conjectures are then confronted with the results of large scale numerical calculations for different forms
of the contour. These results involve the central charge of the RBPM, for which we calculate a precise estimate.

The rest of the paper is organized as follows. The model, its solution in the random cluster representation and the
calculation of the excess entropy is presented in Sec.~\ref{sec:model}.
Numerical results for the corner contribution to the excess entropy are presented in Sec.~\ref{sec:numbeta}. Independent
estimates for the central charge of the model through analyzing the finite-size correction of the free-energy is given in
Sec.~\ref{sec:numc}. Finally, Sec.~\ref{sec:disc} contains our conclusions.

\section{Random bond Potts model in the large-$Q$ limit}
\label{sec:model}

We consider the $Q$-state Potts model defined by the Hamiltonian \cite{Wu}
\be
{\cal H}= - \sum_{\langle i,j\rangle} J_{ij}\delta(\sigma_i,\sigma_j),
\label{hamilton}
\ee
in terms of the Potts spin variables $\sigma_i=0$, $1$, $\dots$,$Q-1$.
The $J_{ij}>0$ couplings between nearest neighbor sites are i.i.d. random variables and in the following we restrict ourselves
to the square lattice. 
In the random cluster representation\cite{Fortuin-Kasteleyn} the partition function of the model at $T=1/\beta$ temperature is given by:
\be
{\cal Z}=\sum_{G}Q^{N_{\text{tot}}(G)}\prod_{ij\in G}\left[e^{\beta J_{ij}}-1\right]
\label{Z}
\ee
where the sum runs over all bond configurations $G$, and in $G$ the total number of connected components (clusters) are
denoted by $N_{\text{tot}}(G)$. The mean number of clusters is given by:
\be
\overline{\left\langle N_\mathrm{tot} \right\rangle}=\dfrac{\partial\overline{ \ln Z(Q)}}{\partial \ln  Q}.
\label{N_tot}
\ee
where $\langle \cdots \rangle$ denotes thermal averaging and $\overline{\cdots}$ stands for the average over quenched disorder.

Let us now introduce a subset of bonds, $\Gamma$, and fix all spins on $\Gamma$ (in state $0$, say) but leave the
couplings unchanged. Then the partition function becomes:
\be
{\cal Z}_{\text{fix}}=\sum_{G}Q^{N_{\text{tot}}(G)-N_{\Gamma}(G)}\prod_{ij\in G}\left[e^{\beta J_{ij}}-1\right]
\ee
where $N_{\Gamma}(G)$ denotes the number of clusters which intersect $\Gamma$. Consequently:
\be
\overline{\left\langle N_\mathrm{tot}-N_{\Gamma}\right\rangle}=\dfrac{\partial \overline{\ln Z_{\text{fix}}(Q)}}{\partial \ln  Q}.
\label{N_tot-N_Gamma}
\ee

Now let us take the large-$Q$ limit, in which case the entropy scales as $S\sim\ln Q$, thus it is convenient
to use the reduced entropy: $S'=S/\ln Q$ and the reduced temperature: $T'=T \ln Q$ ($\beta'=\beta/\ln Q$). These
reduced quantities are of ${\cal O}(1)$ at the phase transition region.

In terms of $\beta'$ the partition function in Eq.(\ref{Z}) reads as
\be
{\cal Z}=\sum_{G}Q^{N_{\text{tot}}(G)}\prod_{ij\in G}\left[Q^{\beta' J_{ij}}-1\right]
\ee
in which for large-$Q$ we have $Q^{\beta' J_{ij}}\gg 1$, thus 
\be
{\cal Z}=\sum_{G}Q^{\phi(G)}\;, \quad \phi(G)=N_{\text{tot}}(G)+\beta'\sum_{ij\in G}J_{ij}\;,
\ee
which is dominated by the largest term, $\phi^*=\max_G\phi(G)$. Finally we arrive at 
\be
{\cal Z}=n_0Q^{\phi^*}\;,
\label{Zc}
\ee
where the degeneracy of the optimal set is $n_0=\mathcal{O}(1)$.
The free-energy of the system, $F$ is proportional to the mean value of $\phi^*$:
\be
\overline{\phi^*}=-\beta' F= S'-\beta'E
\label{free-energy}
\ee
where the reduced entropy of the system is $S'=\overline{N_{\text{tot}}(G^*)}$ and the mean energy is given by:
$E=-\overline{\sum_{ij\in G^*}J_{ij}}$.
Similarly we obtain for the reduced entropy of the system with a contour of fixed spins as:
$S_{\text{fix}}'=\overline{N_{\text{tot}}(G^*)-N_{\Gamma}(G^*)}$,
consequently the excess entropy associated with the contour is given by the mean number of clusters crossed by $\Gamma$:
\be
S'_{\Gamma}=\overline{N_{\Gamma}}\;.
\label{S_N}
\ee
Using this relation $S'_{\Gamma}$ can be calculated numerically, which will be performed in the following section.

On the other hand analytical results on $S'_{\Gamma}$ can be obtained from the difference of
$\overline{\left\langle N_\mathrm{tot} \right\rangle}$ and
$\overline{\left\langle N_\mathrm{tot}-N_{\Gamma}\right\rangle}$ in Eqs.(\ref{N_tot}) and (\ref{N_tot-N_Gamma}),
which is the derivative of $\overline{ \ln Z(Q)}-\overline{\ln Z_{\text{fix}}(Q)}$. At the critical point this
difference is given by: $\sim L_{\Gamma} f_s(Q)+C_{\Gamma} \ln L_{\Gamma}$, where $L_{\Gamma}$ is the linear size
of the contour, $f_s(Q)$ is the surface free-energy density, which is non-universal and the second term represents
the corner contribution\cite{cardy_peschel}. Thus we obtain for the excess entropy:
\be
S'_{\Gamma}=-Qf'_s(Q) L_{\Gamma}+b_{\Gamma}(Q)\ln L_{\Gamma}\;,
\ee
in which the first term corresponds to the ``area-law'' and in the second term the prefactor is factorized as:
\be
b_{\Gamma}(Q)=\frac{\partial c(Q)}{\partial\ln{Q}}A_{\Gamma}=c'A_{\Gamma}\;,
\label{b}
\ee
where $c(Q)=c' \ln Q +{\rm cst.}$ is the central charge of the RBPM for large $Q$. $A_{\Gamma}$ is a geometrical factor,
which does not depend on $Q$ and it follows from the Cardy-Peschel formula\cite{cardy_peschel}
\beqn
A_{\Gamma}=\dfrac{1}{24} \sum_k{ \left( \dfrac{\gamma_k}{\pi}-\dfrac{\pi}{\gamma_k} +\dfrac{2 \pi-\gamma_k}{\pi}-\dfrac{\pi}{2 \pi-\gamma_k}\right)}\;,
\label{A}
\eeqn
where $\gamma_k$ is the interior angle at each corner.

\section{Numerical results for the excess entropy}
\label{sec:numbeta}
In the numerical calculation we have considered finite samples of size $L \times L$ with periodic boundary
conditions and the couplings were taken from a bimodal distribution: 
$P(\beta'J)=\left[\delta(w+\Delta w -\beta'J)+ \delta(w-\Delta w -\beta'J)\right]/2$. Having $w=1/2$ the critical point
is given from self-duality\cite{kinzel}: $\beta'_c=1$. In most of the calculations we have used $\Delta w=1/3$, but to check universality
we have also performed some calculations with $\Delta w=1/4$. The linear size of the systems were $L=32,64,128,256$ and $512$,
and the number of independent samples varied from $80000$ at the smallest sizes to more than $1200$ for $L=512$.

\begin{figure}[!ht]
\begin{center}
\includegraphics[width=3in,angle=0]{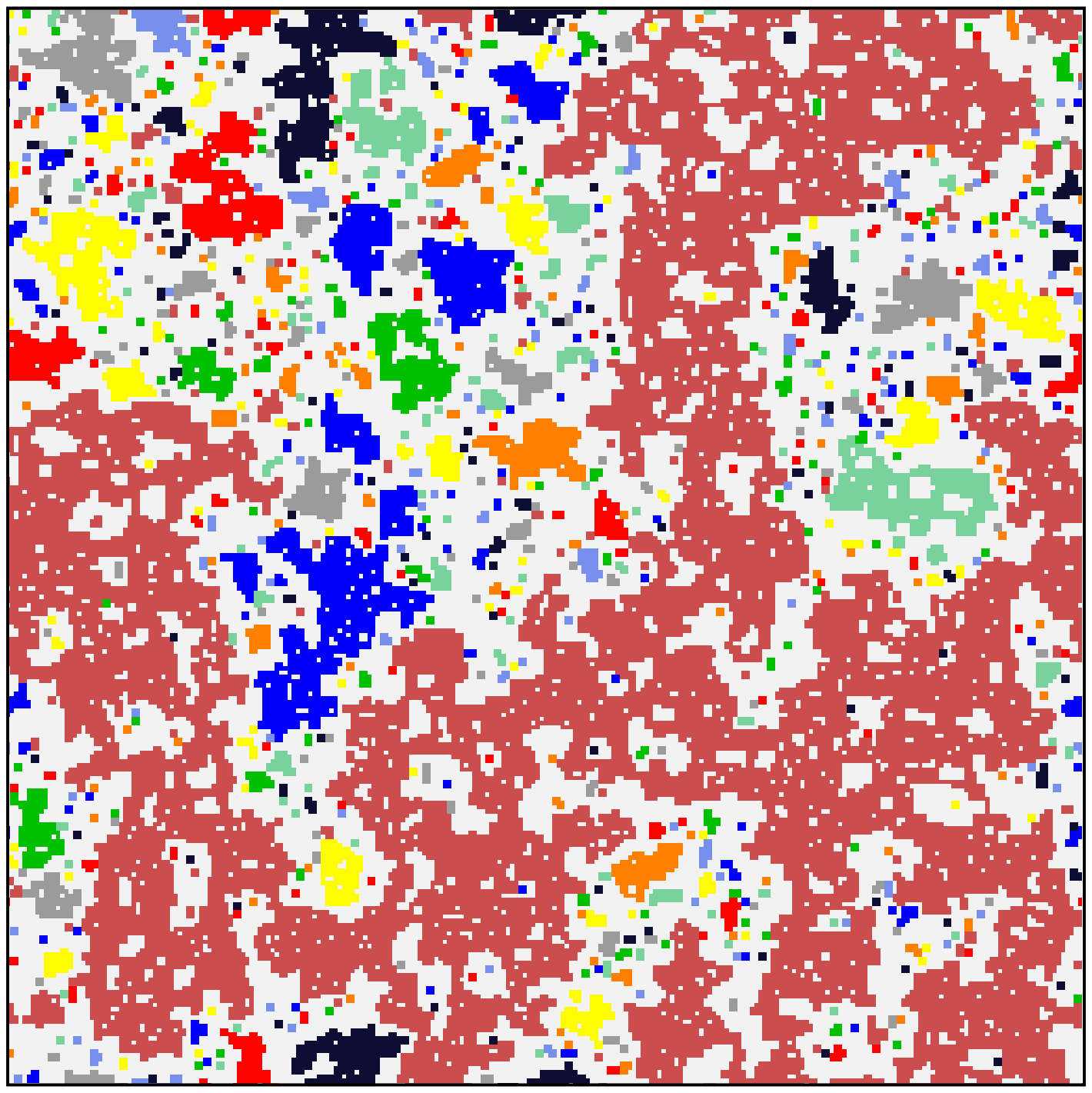}
\includegraphics[width=3in,angle=0]{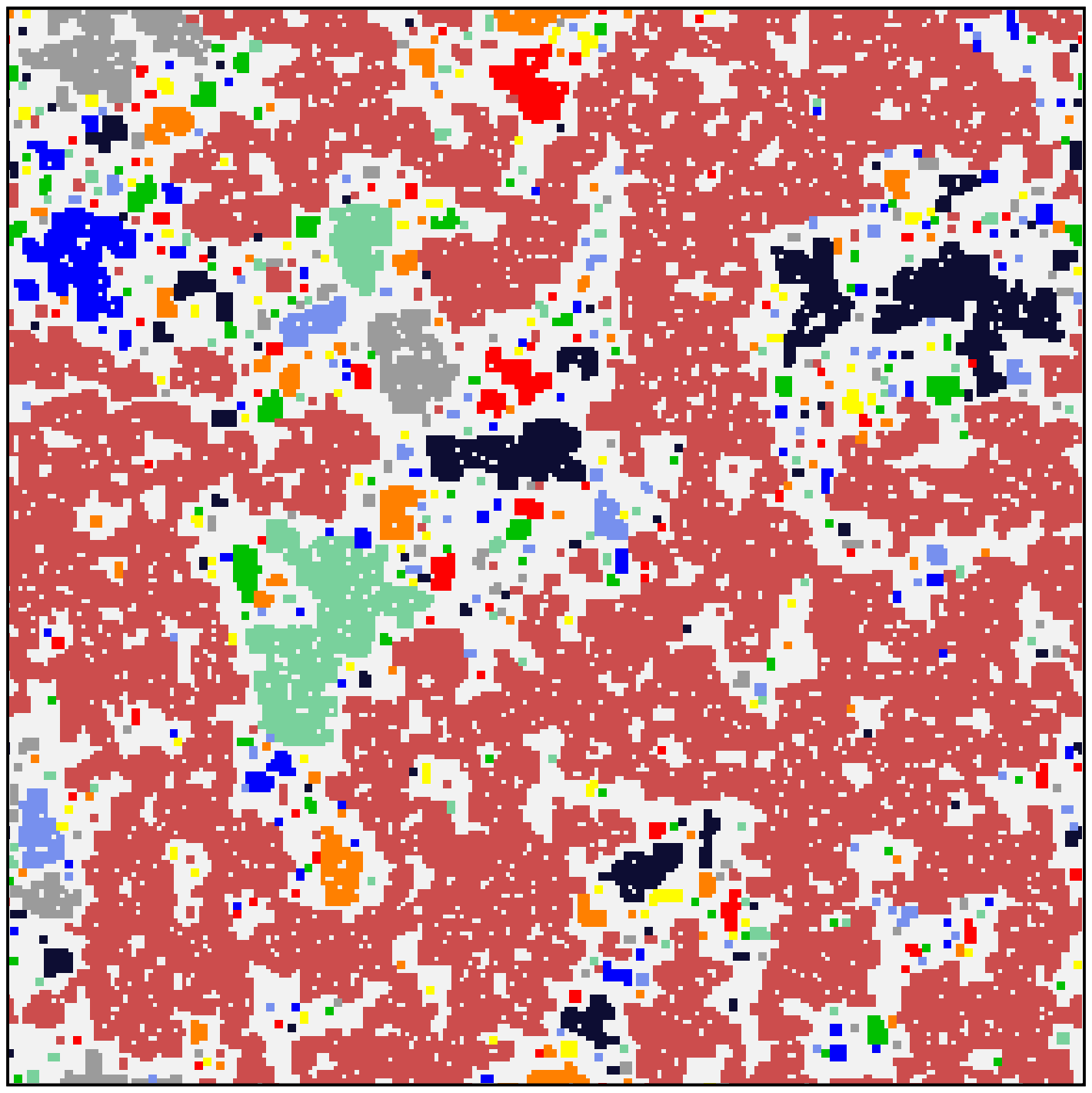}
\end{center}
%\vskip -.5cm
\caption{
\label{fig_1} (Color online) Cluster structure of the same sample in the optimal sets for the intersection (upper panel) and
for the union (lower panel) at $L=256$.}
\end{figure}

The optimal set of a given sample has been calculated by the optimal cooperation algorithm\cite{dips}, which works in polynomial time
and has already been used to solve the RBPM in two-\cite{di2003,mdi2004} and three-dimensional\cite{mercaldo} lattices,
as well as in scale free networks\cite{karsai}.
If there are multiple optimal sets in the system, then both the intersection and union of any pair of them yields an optimal set.
In order to show that our results are independent of the choice of the representing optimal set in a given system,
all our studies are carried out for the two limiting cases, namely for the \emph{union} and \emph{intersection} of \emph{all}
the optimal sets. We illustrate the cluster structure of the optimal sets in these limiting cases in Fig.\ref{fig_1}.
It is seen, that the intersection consists of a large number of smaller clusters, which are partially merged to
common clusters having larger masses in the union. Consequently we have for the averaged number of crossing clusters:
$\overline{N}_{\Gamma}({\rm intersection}) \ge \overline{N}_{\Gamma}({\rm union})$, however the corner contributions being
dominated by the large clusters are expected to be asymptotically identical in the two cases.
%%%%%%%%%% FIG 1  %%%%%%%%%%%%%%%%%%%%%%%%%%%%%%%
%%%%%%%%%% FIG 1  %%%%%%%%%%%%%%%%%%%%%%%%%%%%%%%

\begin{figure}[!ht]
\begin{center}
\includegraphics[width=2.5in,angle=0]{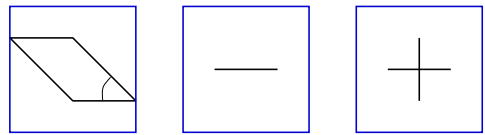}
\end{center}
\vskip -.5cm
\caption{
\label{fig_2} (Color online) Shape of the subsystems used in the numerical calculations: sheared squares, line segments and crosses. 
}
\end{figure}

Having the optimal sets of different samples we have calculated the excess entropy
for different contours: sheared squares, line segments and crosses, which are illustrated in Fig. \ref{fig_2}. To subtract the corner contribution
from the data we have used the so called geometric approach\cite{kovacs_igloi12,kovacs,kovacs14}:
for each sample $N_{\Gamma}$ is calculated in two different geometries,
which have the same boundary term, but different corner ones. Thus the corner contribution is obtained from their difference.
The average in Eq.(\ref{S_N}) is performed over i) different samples and ii) over different ($\sim 1000$) positions of the contour
in a given sample. For technical details we refer to our previous investigations on the non-random model\cite{kovacs,kovacs14,kovacs_3d}.

%%%%%%%%%% FIG 2  %%%%%%%%%%%%%%%%%%%%%%%%%%%%%%%
%%%%%%%%%% FIG 2  %%%%%%%%%%%%%%%%%%%%%%%%%%%%%%%

\begin{figure}[!ht]
\begin{center}
\includegraphics[width=3.4in,angle=0]{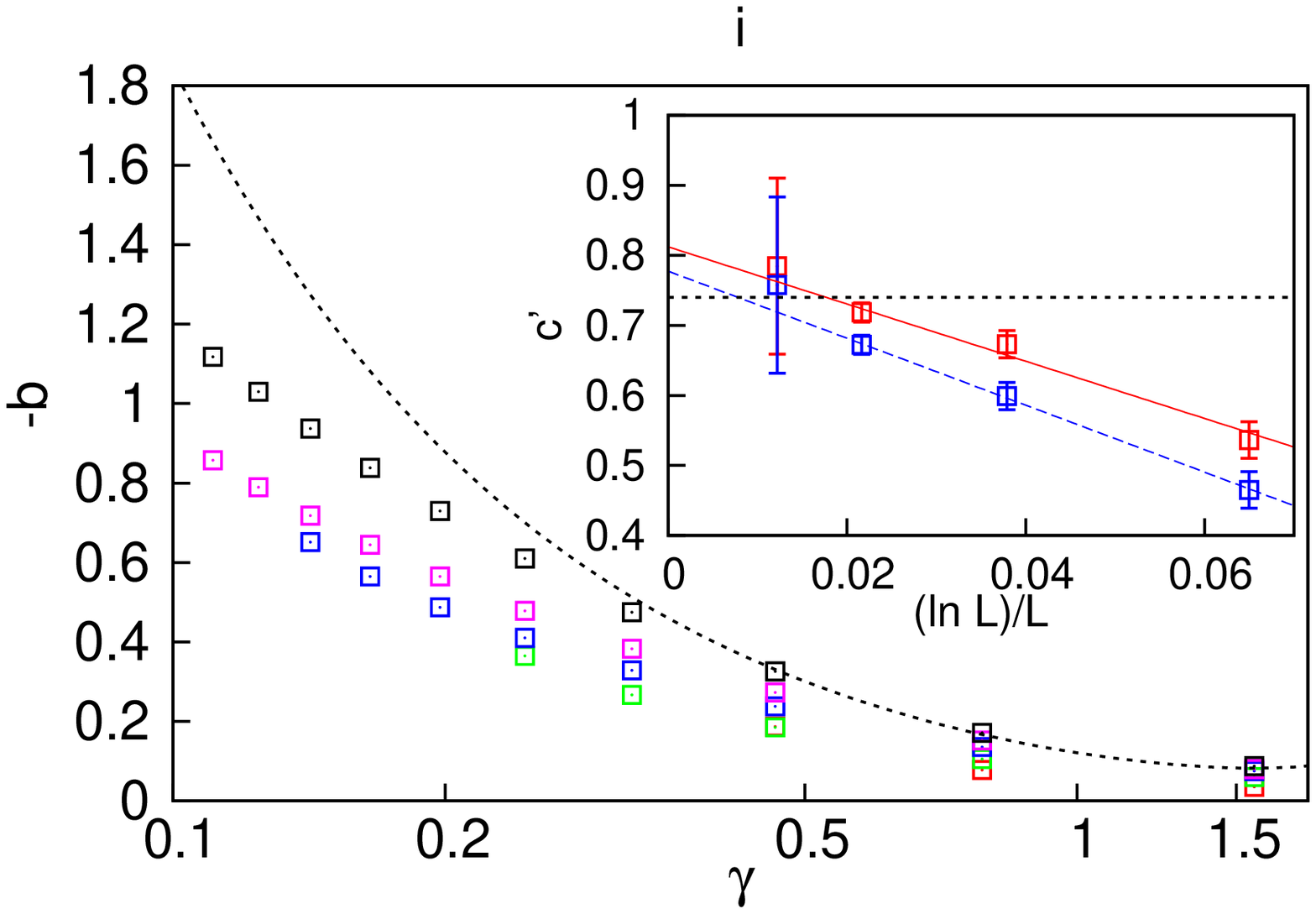}
\includegraphics[width=3.4in,angle=0]{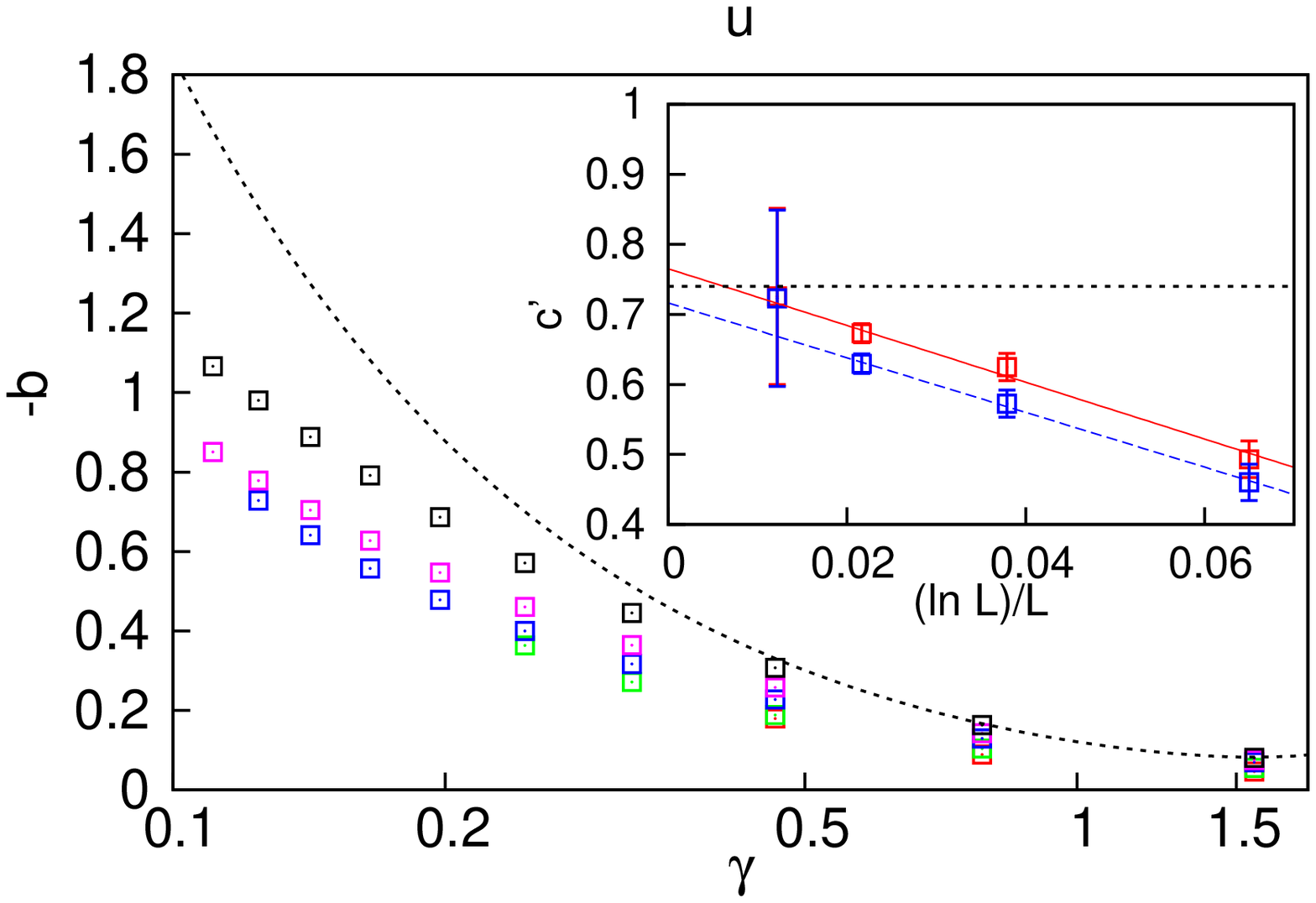}
\end{center}
%\vskip -.5cm
\caption{
\label{fig_3} (Color online) 
Finite size estimates ($L=32$, $64$, $128$, $256$ and $512$, from bottom to top) of the prefactor $b$ with sheared squares as a function of
$\gamma$ for intersection (top) and union (bottom) data. For larger
sizes the results are approaching the conformal result in Eq.(\ref{sheared}), which is indicated by the dotted (black) line.
Inset: Finite size estimates of the reduced central charge $c'$ with sheared squares as a function of the system size
at $\gamma=\pi/2$ ($\gamma=\pi/4$), indicated by solid (dashed) lines. The estimated value $c'=0.74$ is shown by the dotted
horizontal line.}
\end{figure}

We start with \textit{sheared squares}, having an opening angle $\gamma \le \pi/2$ and both its base and altitude is
given by $L/2$. 
In the numerical method we have calculated the corner contribution of the excess entropy for different
sizes and then finite-size estimates are calculated for the prefactor $b_{\Gamma}$ in Eq.(\ref{b}) by two point fit.
These are presented in Fig.\ref{fig_3} for the union and intersection optimal sets. For this contour the geometrical factor,
$A_{\Gamma}$ in Eq.(\ref{A}) reads as
\be
A_{\Gamma}=\frac{1}{12}\left[4-\pi\left(\frac{1}{\gamma}+\frac{1}{\pi-\gamma}+
    \frac{1}{\pi+\gamma}+\frac{1}{2\pi-\gamma}\right)\right]\;,
\label{sheared}
\ee
and we put also the conformal result in Fig.\ref{fig_3} with an estimated reduced central charge $c'=0.74$, see Eq.(\ref{c'}). For not too
small opening angles the finite-size estimates are close to the conformal results, for smaller angles the corrections become larger.
Extrapolating the prefactors for the two largest angles, $\gamma=\pi/2$ and $\pi/4$ gives approximately the same estimate
for the reduced central charge for the intersection and the union, see in Table \ref{table:1}.

%%%%%%%%%% FIG 3  %%%%%%%%%%%%%%%%%%%%%%%%%%%%%%%
%%%%%%%%%% FIG 3  %%%%%%%%%%%%%%%%%%%%%%%%%%%%%%%

\begin{figure}[!ht]
\begin{center}
\includegraphics[width=2.8in,angle=0]{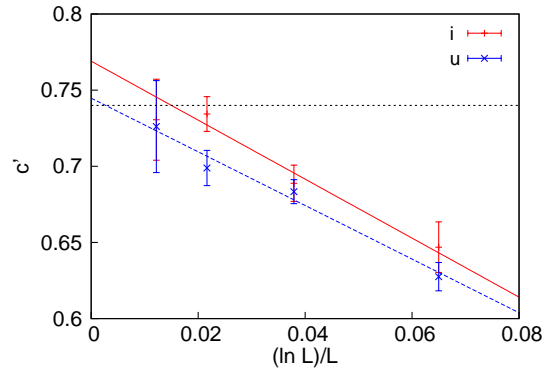}
\end{center}
%\vskip -.5cm
\caption{
\label{fig_4} (Color online) Estimates for the reduced central charge $c'$ for line segments.
The estimated value $c'=0.74$ is indicated by the dotted horizontal line.}
\end{figure}

For \textit{line segments} with length $\ell=L/2$ we have only the corner contributions of two exterior $\gamma=2 \pi$ angles,
so that $A_{\Gamma}=1/8$. In this case the corner contribution is simply half of the number of common clusters between two line segments.
Finite-size estimates for $c'$ are shown in Fig.\ref{fig_4} which are extrapolated to the same value (within the accuracy of the calculation)
both for intersection and union data, see Table \ref{table:1}. We have checked, that the effective central charges for the disorder
parameter, $\Delta w=1/4$, are practically indistinguishable from the results in Fig.\ref{fig_4}.
%
%%%%%%%%%% FIG 4  %%%%%%%%%%%%%%%%%%%%%%%%%%%%%%%
%%%%%%%%%% FIG 4  %%%%%%%%%%%%%%%%%%%%%%%%%%%%%%%
%
%\subsection{Crosses}

\begin{figure}[!ht]
\begin{center}
\includegraphics[width=2.8in,angle=0]{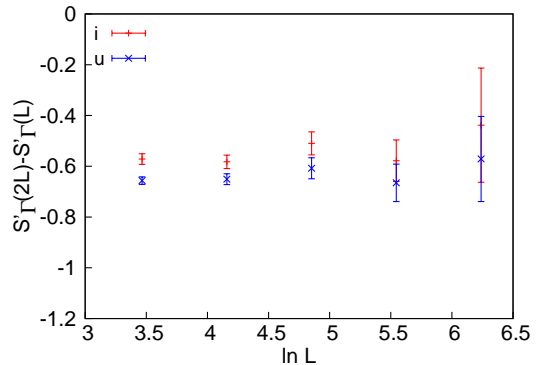}
\end{center}
%\vskip -.5cm
\caption{
\label{fig_5} (Color online) Difference between the excess entropies for one cross with two different sizes: $2L$ and $L$. 
There is no size dependence, hence no corner contribution in agreement
with the Cardy-Peschel formula. 
}
\end{figure}

\begin{figure}[!ht]
\begin{center}
\includegraphics[width=2.8in,angle=0]{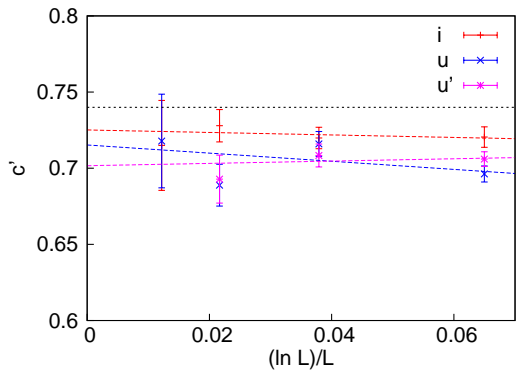}
\end{center}
%\vskip -.5cm
\caption{
\label{fig_6} (Color online) Finite-size estimates for the reduced central charge, $c'$ obtained
by comparing the excess entropies for $1$ and $3$ crosses, for the intersection (i) and the union (u) data.
For the union data we present also the results obtained for the disorder parameter $\Delta w=1/4$ (u').
The estimated value $c'=0.74$ is indicated by the dotted horizontal line.}
\end{figure}

For a \textit{cross-like subsystem} the corner contributions are expected to cancel out completely.
We have checked, that this is indeed the case, see Fig.\ref{fig_5}, showing  the validity of the Cardy-Peschel formula.
We have also studied subsystems comprised of $n=1,2,3$ or $4$ crosses (see in Fig.2 of Ref[\onlinecite{kovacs14}]),
in which case we have obtained finite-size estimates for $c'$. For $1$ and $3$ crosses these are presented
in Fig.\ref{fig_6} and the extrapolated values in Table \ref{table:1}. In this case
results for the disorder parameter $\Delta w=1/4$ leads to the same extrapolated value.
%
%%%%%%%%%% FIG 4  %%%%%%%%%%%%%%%%%%%%%%%%%%%%%%%
%\begin{figure}[!ht]
%\begin{center}
%\includegraphics[width=3.2in,angle=0]{crosses}
%\end{center}
%\vskip -.5cm
%\caption{
%\label{fig_4} (Color online) Subsystems comprising different number of crosses. The value of the geometrical factor, $A_{\Gamma}$
%in Eq. (\ref{A}) is given for the geometries from left to right: $0$, $0$, $-1/8$, $-1/6$ and $-1/4$, which
%is normalized to the area of one cross.
%}
%\end{figure}
%%%%%%%%%% FIG 4  %%%%%%%%%%%%%%%%%%%%%%%%%%%%%%%
%
%%%%%%%%%% FIG 5  %%%%%%%%%%%%%%%%%%%%%%%%%%%%%%%
%%%%%%%%%% FIG 5  %%%%%%%%%%%%%%%%%%%%%%%%%%%%%%%

%%%%%%%%%% FIG 6  %%%%%%%%%%%%%%%%%%%%%%%%%%%%%%%
%%%%%%%%%% FIG 6  %%%%%%%%%%%%%%%%%%%%%%%%%%%%%%%

\section{Central charge from the finite-size correction to the free-energy}
\label{sec:numc}

\begin{figure}[!ht]
\begin{center}
\includegraphics[width=2.8in,angle=0]{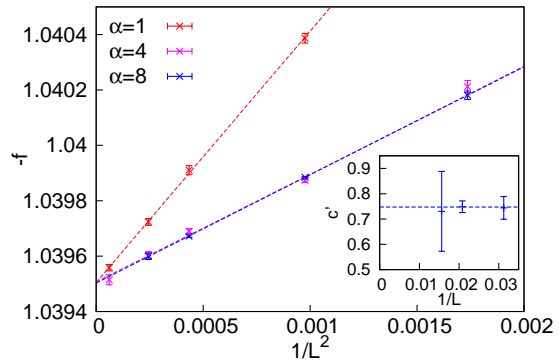}
\end{center}
\vskip -.5cm
\caption{
\label{fig_7} (Color online) Critical free-energy density of $L \times \alpha L$ systems with periodic boundary conditions
as a function of $1/L^2$ for $\alpha=1$ (square) and for $\alpha=4$ and $\alpha=8$ (stripes). The asymptotic value, $f_0=-1.03951(1)$,
is independent of $\alpha$. The prefactor of the correction term is $0.91(3)$ for squares and it agrees for both type of stripes.
In the inset the effective reduced central charge is shown for $\alpha=8$ and the dashed horizontal line represents the estimated
value in Table \ref{table:1}.}

\end{figure}

The traditional way of subtracting the central charge of a two-dimensional lattice model is to study the finite-size correction to the
critical free-energy density in the strip geometry\cite{bloete,affleck}. Having the critical RBPM in an infinite strip of width $L$ with periodic
boundary conditions the free-energy density scales as:
\be
\beta' f(L)=\beta' f_0+\frac{\pi c'}{6L^2}+{\cal O}(L^{-4})\;,
\label{f}
\ee
where $f_0 \equiv f(L=\infty)$ and $c'$ is the reduced central charge in the large-$Q$ limit. Since the free-energy is the same for all optimal sets, we do not make a distinction here between the union and intersection data. In practical adaptation of this method
we have used finite stripes of size $L \times \alpha L$ for $L=24,32,48,64,\dots$ and at a fixed $\alpha \ge 1$ we have plotted the free-energy densities
as a function of $1/L^2$, see Fig. \ref{fig_7}. As expected, the limiting value, $f_0$, does not depend on the aspect ratio, $\alpha$,
but the slopes are different for $\alpha=1$ and for $\alpha=4$ and $8$. For the latter two values the slopes are very close, therefore
we used the data at $\alpha=8$ to estimate effective finite-size reduced central charge, $c'(L)$ from two-point fits.
%comparing $f(L)$ and $f(2L)$.
These are shown in the inset of Fig. \ref{fig_7}. As expected the effective $c'(L)$-s have no noticeable size-dependence for $L \ge 32$,
since the correction terms are ${\cal O}(1/L^2)$. Therefore the extrapolated value given in Table \ref{table:1} is simply the mean
value of the estimates in the inset of Fig. \ref{fig_7}.
%%%%%%%%%% FIG 7  %%%%%%%%%%%%%%%%%%%%%%%%%%%%%%%
%%%%%%%%%% FIG 7  %%%%%%%%%%%%%%%%%%%%%%%%%%%%%%%

\section{Conclusion}
\label{sec:disc}

%%%%%%%%%% TABLE 1  %%%%%%%%%%%%%%%%%%%%%%%%%%%%%%%
\begin{table}[tb]
\caption{Numerical estimates for the reduced central charge $c'$
  from the corner contribution of the excess entropy using different
contour geometries and from
the finite-size correction to the free-energy in stripes.
\label{table:1}}
\begin{tabular}{c|c|c|c|}
 %\hline
 \multicolumn{2}{c|}{} & union & intersection \\ \hline 
 \multirow{3}{*}{\parbox[c]{1.5 cm}{excess\\entropy}} & squares&$ 0.74(3) $
 & $0.79(6)$ \\ \cline{2-4}
 &lines &$ 0.74(3) $& $0.77(5)$ \\ \cline{2-4}
 &crosses &$ 0.72(2) $&$0.73(1)$ \\ \hline 
 \multicolumn{2}{c|}{free-energy} & \multicolumn{2}{|c|}{$ 0.75(2) $}
\end{tabular}
\end{table}
%%%%%%%%%% TABLE 1  %%%%%%%%%%%%%%%%%%%%%%%%%%%%%%%

%%%%%%%%%% FIG 7  %%%%%%%%%%%%%%%%%%%%%%%%%%%%%%%
%\begin{figure}[!ht]
%\begin{center}
%\includegraphics[width=2.8in,angle=0]{Fig_picco}
%\end{center}
%\vskip -.5cm
%\caption{
%\label{fig_picco} (Color online) From the data in Ref.~\onlinecite{picco_e}, the reduced central charge shows a correction proportional
%to $1/\ln{Q}$, with a prefactor of $c'=0.72(3)$. The green (dashed) curve corresponds to a fit comprising also a subleading term, 
%$1/\ln^2{Q}$.
%}
%\end{figure}
%%%%%%%%%% FIG 7  %%%%%%%%%%%%%%%%%%%%%%%%%%%%%%%

The random bond Potts model is a basic problem of statistical physics of disordered systems and its large-$Q$ limit
is of special interest, when the free energy of the system is dominated by a single diagram, the optimal set.
Thermal fluctuations of the $Q \to \infty$ model are negligible compared to disorder fluctuations, therefore its critical properties are
controlled by an infinite disorder fixed point. In this paper we considered this model
on the square lattice and studied the properties of the excess entropy, $S_{\Gamma}$, associated to a contour of bonds,
$\Gamma$. $S_{\Gamma}$ is found to be
proportional to the number of clusters in the optimal set which are crossed by the contour. Using conformal field theory
the excess entropy is shown to have a universal corner contribution at the critical point, which scales with the
logarithm of the linear size of the contour and its prefactor is proportional to the central charge of the model.
We have performed large-scale numerical calculations and confirmed the validity of the conformal prediction. We have
also obtained estimates for the central charge of the model, which are collected in Table \ref{table:1} for different
forms of the contours, as well as for the two extreme forms of the optimal sets. Within the error of the calculation
these all agree with each other, as well as with the results of an independent estimate calculated from the finite-size
dependence of the free-energy density. Based on these data we conclude with the estimate:
\be
c'=0.74(2)\;,
\label{c'}
\ee
for the reduced central charge of the model. This value is close to (although somewhat larger than) the previous estimate\cite{jacobsen}:
$c'=0.72$ obtained through transfer matrix calculations for large, but finite values of $Q$.
$c'$ in Eq.(\ref{c'}) is expected to be universal for any form of the quenched disorder, for the bimodal distribution it holds
for $0<\Delta<1/2$. It is different from that at $\Delta=1/2$, which corresponds to bond percolation having a reduced central
charge\cite{kovacs}
\be
c'_{\text{perc}}=\frac{5\sqrt{3}}{4\pi}\approx0.689.
\label{c}
\ee
It remains the subject
of other research to study if the central charge of the model is related to some properties of the exactly solved
random transverse-field Ising chain\cite{fisher}, as expected from the numerical values of the critical exponents in the two models.

\begin{acknowledgments}
  This work has been supported by the Hungarian National Research Fund under grant No
  OTKA K75324 and K109577. The research of IAK was supported by the European Union and
  the State of Hungary, co-financed by the European Social Fund in the framework of
  T\'AMOP 4.2.4. A/2-11-1-2012-0001 'National Excellence Program'. J.-Ch. A.d'A thanks
  to the 'Theoretical Physics Workshop' and F.I. to the Universit\'e Joseph Fourier
  for supporting their visits to Budapest and Grenoble, respectively.
\end{acknowledgments}


\begin{thebibliography}{99}
\vskip -.5cm

%kell:

\bibitem{shannon} C. Shannon, Bell System Technical Journal \textbf{27}, 379 (1948).

\bibitem{entanglement_entropy} \textit{Entanglement entropy in extended quantum systems}, vol. \textbf{424} of
J. Phys. A (special issue) (2009).

\bibitem{holzhey} C. Holzhey, F. Larsen, and F. Wilczek, Nucl.Phys. B \textbf{424}, 443 (1994).

\bibitem{vidal} G. Vidal, J. I. Latorre, E. Rico, and A. Kitaev, Phys.Rev.Lett. \textbf{90}, 227902 (2003).

\bibitem{calabrese_cardy} P. Calabrese and J. Cardy, J. Stat. Mech. P06002 (2004).

\bibitem{wilms} J. Wilms, M. Troyer, and F. Verstraete, J. Stat. Mech. P10011 (2011).

\bibitem{lau_grassberger}H. W. Lau, and P. Grassberger, Phys Rev E \textbf{87}, 022128 (2013).

\bibitem{mutual_classic} J.-M. St\'ephan, S. Inglis, P. Fendley, R. G. Melko, Phys. Rev. Lett. \textbf{112}, 127204 (2014)

\bibitem{Wu}
  F.Y. Wu, Rev. Mod. Phys. {\bf 54}, 235 (1982).
  
\bibitem{Cardy99}
        For a review, see: J.L. Cardy, Physica A{\bf 263}, 215 (1999).

\bibitem{imry} Y. Imry and M. Wortis, Phys. Rev. B \textbf{19}, 3580 (1979).

\bibitem{aizenman_wehr} M. Aizenman and J. Wehr, Phys. Rev. Lett. \textbf{62}, 2503 (1989); \textbf{64}, 1311(E) (1990).

\bibitem{hui} K. Hui and A. N. Berker, Phys. Rev. Lett. \textbf{62}, 2507 (1989); \textbf{63}, E2433 (1989).

\bibitem{cardyj} J. Cardy and J. L. Jacobsen, Phys. Rev. Lett. \textbf{79}, 4063 (1997).

\bibitem{jcardy} J. L. Jacobsen and J. Cardy, Nucl. Phys. B \textbf{515} [FS], 701(1998).

\bibitem{picco} M. Picco, Phys. Rev. Lett. \textbf{79}, 2998 (1997); C. Chatelain
and B. Berche, Phys. Rev. Lett. \textbf{80}, 1670 (1998); Phys.
Rev. E \textbf{58} R6899 (1998); \textbf{60}, 3853 (1999); T. Olson and
A.P. Young, Phys. Rev. B \textbf{60}, 3428 (1999).

\bibitem{palagyi} G. Pal\'agyi, C. Chatelain, B. Berche, and F. Igl\'oi, Eur. Phys. J. B{\bf 13}, 357 (2000).

\bibitem{chatelain} C. Chatelain, EPL \textbf{102},66007 (2013); Phys. Rev. E \textbf{89}, 032105 (2014).

\bibitem{baxter} R. J. Baxter, J. Phys. C \textbf{6}, L445 (1973).

\bibitem{jacobsen} J.L. Jacobsen and M. Picco, Phys. Rev. E \textbf{61}, R13 (2000);
M. Picco (unpublished).

\bibitem{juhasz} R. Juh\'asz, H. Rieger, and F. Igl\'oi, Phys. Rev. E\textbf{64}, 056122 (2001).

\bibitem{Fortuin-Kasteleyn} P. W. Kasteleyn and C. M. Fortuin, J. Phys. Soc. Japan \textbf{26}, 11 (1969).

\bibitem{dips} J.-Ch. Angl\`es d'Auriac, F. Igl\'oi, M. Preissmann, and A.
Seb\H{o}, J. Phys. A \textbf{35}, 6973 (2002); J.-Ch. Angl\`es d'Auriac, in {\it New Optimization Algorithms in Physics},
        edt. A.~K. Hartmann and H. Rieger (Wiley-VCH, Berlin 2004).

\bibitem{di2003} J.-Ch. Angl\`es d'Auriac and F. Igl\'oi, Phys. Rev. Lett. \textbf{90}, 190601 (2003).

\bibitem{mdi2004} M.T. Mercaldo, J.-Ch. Angl\`es d'Auriac and F. Igl\'oi, Phys. Rev. E \textbf{69}, 056112 (2004).

\bibitem{im}
       For a review, see: F. Igl\'oi, and C. Monthus, Phys. Rep. {\bf 412}, 277 (2005).
        
\bibitem{fisher} D.S. Fisher, Phys. Rev. Lett. 69, 534 (1992); Phys. Rev. B \textbf{51}, 6411 (1995).

\bibitem{kovacs} I. A. Kov\'acs, F. Igl\'oi and J. Cardy, Phys. Rev. B \textbf{86}, 214203 (2012).

\bibitem{kovacs_3d} I. A. Kov\'acs, and F. Igl\'oi, Phys. Rev. B \textbf{89}, 174202 (2014).

\bibitem{kovacs14} I. A. Kov\'acs, E. M. El\c{c}i, M. Weigel and F. Igl\'oi,
Phys. Rev. B\textbf{89}, 064421 (2014).

\bibitem{cardy_peschel} J. Cardy and I. Peschel, Nucl. Phys. B \textbf{300} [FS22], 377 (1988).

\bibitem{kinzel} W. Kinzel and E. Domany, Phys. Rev. B \textbf{23}, 3421 (1981).

\bibitem{mercaldo} M. T. Mercaldo, J-Ch. Angl\`es d'Auriac, and F. Igl\'oi, Europhys. Lett. {\bf 70}, 733 (2005);
Phys. Rev. E{\bf 73}, 026126 (2006).

\bibitem{karsai} M. Karsai, J-Ch. Angl\`es d'Auriac, and F. Igl\'oi, Phys. Rev. E {\bf 76}, 041107 (2007).

\bibitem{kovacs_igloi12} I. A. Kov\'acs and F. Igl\'oi, EPL \textbf{97}, 67009 (2012).

\bibitem{bloete} H. W. J. Bloete, J. L. Cardy, and M. P. Nightingale, Phys. Rev. Lett. \textbf{56}, 742 (1986).

\bibitem{affleck} I. Affleck, Phys. Rev. Lett. \textbf{56}, 746 (1986).

%\bibitem{cardy_conf} J.L. Cardy, in \emph{Phase Transitions and Critical Phenomena}, edited by C. Domb
%and J.L. Lebowitz (Academic Press, London, 1987), Vol. 11. p. 55.
%
%\bibitem{stauffer} D. Stauffer and A. Aharony, \textit{Introduction to percolation theory} (2nd ed.), Taylor \& Francis, London, 1992.
%
%\bibitem{yu07} R. Yu, H. Saleur and S. Haas, \prb \textbf{77}, 140402 (2008).
%
%\bibitem{area} J. Cardy and R.M. Ziff, J. Stat. Phys. \textbf{110}, 1 (2003).
%
%\bibitem{zamolodchikov} A. B. Zamolodchikov, Pis'ma Zh. Eksp. Teor. Fiz. 43, 565 (1986) [JETP Lett. 43, 730 (1986)].
%
%\bibitem{ds} J. C. Angl\`es d'Auriac and Nicolas Sourlas, Europhys. Lett. 39, 473 (1997).
%
%
%\bibitem{chen} S. Chen, A. M. Ferrenberg, and D. P. Landau, Phys. Rev. E 52, 1377 (1995).
%
%\bibitem{picco_e} M. Picco, Phys. Rev. E \textbf{61}, R13 (2000).
%
%\bibitem{olson} T. Olson and A. P. Young, Phys. Rev. B 60, 3428 (1999).

%di2003:

%\bibitem{ccfs} J. T. Chayes, L. Chayes, D. S. Fisher and T. Spencer,
%Phys. Rev. Lett. 57, 299 (1986).
%\bibitem{kogut} J. Kogut, Rev. Mod. Phys. 51, 659 (1979).
%[1] B.M. McCoy and T.T. Wu, Phys. Rev. 176, 631 (1968); 188, 982(1969); B.M. McCoy, Phys. Rev. 188, 1014 (1969).
%[2] O. Motrunich, S.-C. Mau, D.A. Huse and D.S. Fisher, Phys. Rev. B61, 1160 (2000).
%[3] D.S. Fisher, Phys. Rev. Lett. 69, 534 (1992); Phys. Rev. B 51, 6411 (1995).
%[5] F. Igl´oi and H. Rieger, Phys. Rev. Lett. 78, 2473 (1997).
%[7] F.Y. Wu, Rev. Mod. Phys. 54, 235 (1982).
%[17] S.K. Ma, C. Dasgupta and C.-K. Hu, Phys. Rev. Lett. 43, 1434 (1979); C. Dasgupta and S.K. Ma, Phys. Rev. B22, 1305 (1980).

%jp-bol:

%\bibitem{ludwigc} A. W. W. Ludwig and J. L. Cardy, Nucl. Phys. B 285 [FS19], 687 (1987).
%\bibitem{ludwig} A. W. W. Ludwig, Nucl. Phys. B 285, 97 (1987); 330, 639 (1990).
%\bibitem{dotsenko} Vl. S. Dotsenko, J. L. Jacobsen, M.-A. Lewis, and M. Picco, Nucl. Phys. B 546 [FS], 505 (1999).
%\bibitem{baxter} R. J. Baxter, Exactly Solved Models in Statistical Mechanics (Academic Press, London, 1982).
%\bibitem{bn} H. W. J. Bl\"ote and M. P. Nightingale, Physica A 112, 405 (1982).
%\bibitem{bcn} H. W. Bl\"ote, J. L. Cardy, and M. P. Nightingale, Phys. Rev. Lett. 56, 742 (1986); I. Affleck, ibid. 56, 746 (1986).
%\bibitem{furstenberg} H. Furstenberg, Trans. Am. Math. Soc. 68, 377 (1963).

%kovacs_potts-bol:

%\bibitem{potts} R.B. Potts, Proc. Cambridge Philos. Soc. 48, 106 (1952).
%\bibitem{wu} F.Y. Wu, Rev. Mod. Phys. 54, 235 (1982).
%\bibitem{baxter} R.J. Baxter, J. Phys. C \textbf{6}, L445 (1973).
%R.J. Baxter, \textit{Exactly solved models of statistical mechanics}, Academic Press, London, 1982.
%\bibitem{nienhuis-review} B. Nienhuis in \emph{Phase Transitions and Critical Phenomena}, edited by C. Domb and J.L. Lebowitz (Academic Press, London, 1987), Vol. 11. p. 1.
%\bibitem{sle} O. Schramm, Israel J. Math. 118, 221 (2000); S. Smirnov and W. Werner, Math. Research Letters 8, no. 5-6, 729-744 (2001).
%\bibitem{coniglio_klein} A. Coniglio and W. Klein, J. Phys. A \textbf{13}, 2775 (1980).
%\bibitem{hu} C. K. Hu, Phys. Rev. B 29, 5103 (1984).
%\bibitem{dotsenko_fateev} Vl.S. Dotsenko and V.A. Fateev, Nucl. Phys. B \textbf{240}, 312 (1984).
%\bibitem{deng} Y. Deng, H.W.J. Bl\"ote and B. Nienhuis, Phys. Rev. E \textbf{69}, 026123 (2004).
%\bibitem{janke-geo} W. Janke and A.M.J. Schakel, Nucl. Phys. B \textbf{700}, 385 (2004).
%\bibitem{duplantier} B. Duplantier, J. Stat. Phys. 110, 691 (2003).
%\bibitem{nbrs} B. Nienhuis, A.N. Berker, E.K. Riedel and M. Schick, Phys. Rev. Lett. 43, 737 (1979).
%\bibitem{stella_vanderzande} A.L. Stella and C. Vanderzande, Phys. Rev. Lett. \textbf{62}, 1067 (1989).
%\bibitem{duplantier_saleur} B. Duplantier and H. Saleur, Phys. Rev. Lett. \textbf{63}, 2536 (1989).
%\bibitem{qian} X. Qian, Y. Deng and H. W. J. Bl\"ote, Phys. Rev. E {\bf 72}, 056132 (2005).
%\bibitem{vanderzande} C. Vanderzande, J. Phys. A \textbf{25}, L75 (1992).
%\bibitem{weigel} M. Weigel and W. Janke, Phys. Lett. B \textbf{639}, 373 (2006).
%\bibitem{dubail1} J. Dubail, J.L. Jacobsen and H. Saleur, J. Phys. A \textbf{43}, 482002 (2010).
%\bibitem{dubail2} J. Dubail, J.L. Jacobsen and H. Saleur, arXiv:1010.1700.
%\bibitem{threepoint} G. Delfino, M. Picco, R. Santachiara and J. Viti, arXiv:1307.6123.
%\bibitem{murata} K.K. Murata, J. Phys. A \textbf{12}, 81 (1979).
%\bibitem{coniglio} A. Coniglio and F. Peruggi, J. Phys. A \textbf{15}, 1873 (1982).
%\bibitem{swendsen} R.H. Swendsen and J.S. Wang, Phys. Rev. Lett. 58, 86 (1987).
%\bibitem{eren_martin13} E. M. El\c{c}i and M. Weigel, Phys. Rev. E \textbf{88}, 033303 (2013).
%\bibitem{Q4log} J. L. Cardy, M. Nauenberg, and D. J. Scalapino, Phys. Rev. B 22, 2560 (1980).
%\bibitem{cardy01} J. Cardy, Phys. Rev. Lett., 84, 3507 (2000).
%\bibitem{lin07} Y-C. Lin, F. Igl\'oi and H. Rieger, \prl \textbf{99}, 147202 (2007).

\end{thebibliography}
\end{document}